\documentstyle[12pt,epsf]{article}
 \font\blackboard=msbm10 
 \font\blackboards=msbm7 \font\blackboardss=msbm5
 \newfam\black \textfont\black=\blackboard
 \scriptfont\black=\blackboards \scriptscriptfont\black=\blackboardss
 \def\Bbb#1{{\fam\black\relax#1}}

\font\ninerm=cmr9
\def\uniset{\rlap{\ninerm 1}\kern.15em 1}

\def\d{{\rm d}}
\def\mi{{\rm i}}
\def\j{{\rm j}}
\def\G{\mathop{\Gamma}\nolimits}
\def\Re{\mathop{\rm Re}\nolimits}
\def\e{\mathop{\rm e}\nolimits}
\def\sq2{\sqrt{2}}

\def\defi{\stackrel{\rm def}{=}}
\newcommand\grsim{\mathrel{\hbox{\lower1ex\hbox{\rlap{$\sim$}\raise1ex\hbox{$>$}}}}}
\newcommand\losim{\mathrel{\hbox{\lower1ex\hbox{\rlap{$\sim$}\raise1ex\hbox{$<$}}}}}

\title{Exact resolution method for general 1D polynomial Schr\"odinger equation}
\author{{\bf Andr\'e Voros} \\
\\
CEA--Saclay, Service de Physique Th\'eorique\\
F-91191 Gif-sur-Yvette CEDEX (France)\\
{ E-mail : voros@spht.saclay.cea.fr}\\
\\
and\\
\\
Institut de Math\'ematiques de Jussieu\\
CNRS UMR 7586\\
Universit\'e Paris 7\\
2 place Jussieu,
F-75251 Paris CEDEX 05 (France)}

\begin{document}
\maketitle
{\abstract 
The stationary 1D Schr\"odinger equation with a polynomial potential $V(q)$
of degree $N$ is reduced to a system of exact quantization conditions of
Bohr--Sommerfeld form. 
They arise from bilinear (Wronskian) functional relations pairing
spectral determinants of $(N+2)$ generically distinct operators, 
all the transforms of one quantum Hamiltonian 
under a cyclic group of complex scalings. 
The determinants' zeros 
define $(N+2)$ semi-infinite chains of points in the complex spectral plane,
and they encode the original quantum problem.
Each chain can now be described by an exact quantization condition 
which constrains it in terms of its neighbors,
resulting in closed equilibrium conditions for the global chain system;
these are supplemented by the standard (Bohr--Sommerfeld) quantization
conditions, which bind the infinite tail of each chain asymptotically.
This reduced problem is then probed numerically for effective solvability
upon test cases (mostly, symmetric quartic oscillators): 
we find that the iterative enforcement of all the quantization conditions
generates discrete chain dynamics which appear to converge geometrically 
towards the correct eigenvalues/eigenfunctions. 
We conjecture that the exact quantization then acts by 
specifying reduced chain dynamics which can be stable (contractive) 
and thus determine the exact quantum data as their fixed point.
(To date, this statement is verified only empirically 
and in a vicinity of purely quartic or sextic potentials $V(q)$.)
}

\bigskip

We study the 1-dimensional stationary Schr\"odinger equation 
with a real polynomial potential $V(q)$ of degree $N\ (>2)$ on the real axis,
taken in the rescaled form
\begin{equation}
\label{Schr}
-\psi''(q) + [V(q) + \lambda] \psi(q) = 0, \qquad  
V(q) \equiv +q^N + v_1 q^{N-1} + v_2 q^{N-2} + \cdots + v_{N-1} q.
\end{equation}
We write the collection of coefficients as $\vec v \defi (v_1,\cdots,v_{N-1})$,
absorbing any constant term of $V$ into the spectral parameter $\lambda$
(here the {\sl sign-reverse\/} of the usual energy).

All standard analytical treatments of this problem lead to 
asymptotic results at best, i.e., to semiclassical or perturbative expansions
which diverge factorially and are not even Borel-summable in general
(\cite{BW}--\cite{V4} provide some directly related references).
Even though sophisticated techniques can sometimes convert those results
to converging numerical outputs,
those pathologies signal that such approaches may be mishandling 
the exact analytical structure of the problem (\ref{Schr}).

In this work we further develop an alternative and entirely exact 
method, which until now was specialized to quantizing the spectrum
in the fully homogeneous case $\vec v =0$ \cite{V2,V3,V4}.
There, it produced selfconsistent, Bohr--Sommerfeld-like quantization formulae
whose implementation (through iterative numerical schemes)
apparently reconstructs the exact spectrum
(convergent iterations are always observed, but this fact remain to be proved).
The derivation of those results was quite indirect,
via Borel transformations and quantum resurgence; 
this made calculations lengthy, hard to generalize, 
and pending on still incompletely established regularity properties
of the Borel-transformed solutions (cf. \cite{E}; thm. 1.2.1 in \cite{DPA}).

The approach now to be described towards the same goal is much more direct, 
and immediately applicable to general (inhomogeneous) potentials.
It only proceeds through a pair of auxiliary eigenvalue problems on a half-line,
described in Sec.1; the spectral determinants of these problems exactly express
the wave-function data at the endpoint through the identity (\ref{ID}),
and thus they inherit a bilinear (Wronskian) functional relation, eq.(\ref{DW}),
through an analysis closely following Sibuya \cite{S}.
Exact quantization conditions are then readily extracted 
for the corresponding Dirichlet/Neumann spectra, 
representing our analytical end result: eqs.(\ref{EQC}) (Sec.2).
As in the homogeneous case, these determine the unknown spectra 
only as fixed points of (explicit) iterations; 
the convergence of the latter is then a meaningful separate question,
which here remains conjectural but is numerically probed in the remaining 
Secs.3--4 on several test cases.
First on even potentials, with the previous spectra directly giving 
the eigenvalues of eq.(\ref{Schr}) over the whole line (Sec.3).
Then, because eqs.(\ref{ID}) also supply the wave-function data 
at the endpoint of the half-line, the straight variation of this endpoint 
yields the solution of eq.(\ref{Schr}) itself; 
this allows us to also test the formalism 
on a (ground state) eigenfunction calculation (Sec.4). 

\section{Spectral preliminaries}

\subsection{Polarized boundary conditions}

We first cast eq.(\ref{Schr}) into two eigenvalue problems 
with asymmetric boundary conditions. 
Specifically, we restrict the problem to a half-line where $V$ is confining:
$[0,+\infty)$, we keep the square-integrability at $+\infty$, 
while we put a Neumann, 
resp. Dirichlet, boundary condition at the finite endpoint $q=0$.
Both conditions then define self-adjoint operators $\hat H^+$, resp. $\hat H^-$,
which have purely discrete spectra
we respectively denote $\{E_{2k}\}$ and $\{E_{2k+1}\}$ for $k=0,1,2, \ldots$; 
these admit an asymptotic expansion which is the Bohr--Sommerfeld formula
reexpanded in descending fractional powers of the energy,
\begin{equation}
\label{BS}
\sum_\nu {\tilde b}_\nu E_k^\nu
\sim (k+1/2),\qquad k \to \infty {\rm \ in\ }\Bbb N, \quad 
\nu=\mu,\ \mu-1/N,\ \mu-2/N, \cdots
\end{equation}
\begin{eqnarray}
\label{MU}
{\rm with} \quad \mu \defi {1 \over 2}+ {1 \over N},\qquad 
{\tilde b}_\mu \defi \oint_{p^2+q^N=1} { p \,\d q \over 2\pi } =
{\pi^{-1/2}\over N}\G\Bigl({1 \over N}\Bigr) \Big/ 
\G\Bigl({3 \over 2}+{1 \over N}\Bigr) ,\\
{\rm and} \quad {\tilde b}_{\mu-j/N}(\vec v) = 
{\rm a\ polynomial\ in\ the\ } v_{j'}\ (j' \le j), \ e.g.,\quad 
{\tilde b}_{\mu-1/N} = -{2v_1 \over \pi N}\,;\nonumber
\end{eqnarray}
in complete generality, the series depends on the sector (Neumann vs Dirichlet):
${\tilde b}_\nu = {\tilde b}^\pm_\nu$, but not until $\nu = -3/2$;
whereas we will invoke those coefficients (and related ones, see Sec.1.4) 
only in the leading range $\{\nu >{-\mu}\}$, before this complication appears.
(If $V$ is an even polynomial, the two spectra coincide with the even vs odd
parity components of the spectral problem on the whole line, 
hence their asymptotics coincide to all orders.)

Concerning the exponent $\mu$ (the {\sl growth order\/}),
the forthcoming arguments often assume the property $\mu<1$ (i.e., $N>2$)
and always $\mu\ne 1$: the case $N=2$ (singular) must be ruled out \cite{V4}.

Then, up to sign, the eigenvalues read as the zeros of the Fredholm determinants
\begin{equation}
\label{Fred}
\Delta^\pm (\lambda) \defi 
\prod_{k\ {\scriptstyle{\rm even}\atop \scriptstyle{\rm odd}}}
(1+ \lambda / E_k);
\end{equation}
these infinite products converge to entire functions of order $\mu$ in the variable $\lambda$, also entire in the parameters $\vec v$ \cite{S}.

At fixed $(\lambda,\vec v)$, 
let $\psi_\lambda (q) $ denote a recessive solution of eq.(\ref{Schr}) 
(i.e., a solution exponentially decreasing as $q \to +\infty$, 
which is unique up to normalization).
Then $ \psi_\lambda (0) $ vanishes simultaneously with $ \Delta^- (\lambda)$,
(and likewise for $ \psi'_\lambda (0) $ and $ \Delta^+ (\lambda) $).
However, the relations connecting them retain cumbersome factors unless 
each of $\Delta $ and $ \psi$ is suitably renormalized.
The procedure for the determinants is the well-known zeta-regularization,
but a separate rescaling of the wave functions also contributes in parallel
to optimal simplification.

\subsection{The spectral (or functional) determinants}

By Mellin-transforming eq.(\ref{BS}), the spectral zeta functions 
(in analogy with \cite{VZ})
\begin{equation}
\label{Z} 
Z^\pm (s) \defi 
\sum_{k\ {\scriptstyle{\rm even}\atop \scriptstyle{\rm odd}}} E_k^{-s}
\qquad  ({\rm and} \quad Z(s) \defi Z^+ (s) +Z^- (s) )
\qquad  \qquad (\Re s > \mu)
\end{equation}
are seen to extend {\it meromorphically\/} to lower $\Re s$ values, 
and to be regular at integer points. An effective analytical continuation is
brought by an Euler--Maclaurin summation formula which uses the asymptotic
information (\ref{BS}): the regularized form
\begin{eqnarray}
\label{EML}
Z^\pm (s) = \lim _{K \to +\infty} 
\Bigl\{ \sum_ {k<K} E_k^{-s} +{1\over 2} E_K^{-s}
-{1\over 2} \sum_{\{\nu>\nu_0\}} {\nu{\tilde b}_\nu \over (-s+\nu)} E_K^{-s+\nu}
\Bigr\} \quad
{\rm for\ }k,K\ \scriptstyle{\scriptstyle{\rm even}\atop \scriptstyle{\rm odd}}
\end{eqnarray}
converges as soon as $\nu_0$ can be taken below $(\Re s)$.
At the regular point $s=0$, this gives
\begin{equation}
\label{Z0}
Z^\pm(0 )= {{\tilde b}_0 \over 2} \pm {1\over 4}\ .
\end{equation}

The point $s=0$ further serves to define spectral determinants, as 
{\sl zeta-regularized products\/}: $\det \hat H^\pm \defi \exp [-{Z^\pm}'(0)]$
and, by straightforward extension,
\begin{equation}
\label{DET}
\quad D^\pm(\lambda) \defi \det (\hat H^\pm + \lambda) 
\equiv \exp [- {Z^\pm}'(0)]\, \Delta^\pm (\lambda).
\end{equation}
This has the even more explicit Euler--Maclaurin representation
\begin{eqnarray}
\label{EMc}
\log D^\pm(\lambda)  &=& \lim _{K \to +\infty} \Bigl\{ \sum_ {k<K}
\log (E_{k} + \lambda) + {1\over 2} \log (E_{K} + \lambda) \\
&&\qquad -{1 \over 2}\sum_{\{\nu>0\}} {\tilde b}_\nu E_{K}^\nu
\Bigl( \log E_{K}-{1 \over \nu}\Bigr) \Bigr\} \quad
{\rm for\ }k,K\ \scriptstyle{\scriptstyle{\rm even}\atop \scriptstyle{\rm odd}}.
\nonumber
\end{eqnarray}
Crucially, we can and will form these zeta-regularized products
for general sequences admitting asymptotics of the form
(\ref{BS}), including complex spectra \cite{QHS} in which case 
the formulae are interpreted by analytical continuation 
from $\{\lambda>0, \ \vec v =0\}$. 
The $\{E_k\}$ are the zeros of the zeta-regularized product and they 
{\sl completely specify it\/} (as opposed to a Hadamard infinite product,
for which at least a normalizing factor has to be supplied independently);
this normalization {\sl commutes with global spectral translations\/}
(unlike eq.(\ref{Fred})).

\subsection{Absolute WKB normalization for recessive solutions}

A recessive solution of eq.(\ref{Schr}) on ${\Bbb R}^+$ admits 
an exact WKB representation \cite{V1} (for $\lambda > -\inf_{{\Bbb R}^+} V$)
\begin{equation}
\label{WKB}
 \psi_{\lambda,q_0} (q) \equiv u(q)^{-1/2} \e^{-\int_{q_0}^q u(q') \d q'}
\quad {\rm where,\ for\ } q \to +\infty , \quad
u(q) \sim
\Pi (q) \defi (V(q) + \lambda)^{1/2}.
\end{equation}
This normalization awkwardly depends on the base point $q_0$, 
so we seek a more intrinsic one with $q_0=+\infty$:
\begin{equation}
\label{NN}
\psi_\lambda (q) \defi u(q)^{-1/2} \e^{\int_q^{+\infty} u(q') \d q'}.
\end{equation}
However, only $(u-\Pi)(q)$ is integrable at infinity 
whereas $u(q) \sim \Pi (q) \sim q^{N/2}$, 
so we add a further prescription, 
\begin{equation}
\label{zz}
\int_q^{+\infty} u(q') \d q' \defi 
\int_q^{+\infty} (u - \Pi) \d q' 
+ \int_q^{+\infty} (V(q') + \lambda)^{-s} \d q' |_{s=-1/2}
\end{equation}
where the latter integral is now defined by analytical continuation 
from the domain $\{\Re s > 1/N,\ \lambda > -\inf V \}$.
For the basic example $V(q)=q^N$, in the notations of eq.(\ref{MU}),
\begin{equation}
\int_0^{+\infty} (q^N + \lambda)^{-s} \d q \equiv 
{\G(s-{1 \over N}) \G({1 \over N}) \over N \G(s)} \lambda^{-s+1/N}
\quad \Rightarrow  \quad \int_0^{+\infty} (q^N + \lambda)^{1/2} \d q = 
{\pi {\tilde b}_\mu \over 2 \sin\pi\mu} \lambda ^\mu . 
\end{equation}

In general, however, $\int_q^{+\infty} (V + \lambda)^{-s}$
 develops a singularity at $s=-1/2$,
which will explicitly affect many formulae hence requires further study.
This singularity is immediately seen to be independent of $q$
(since $\int_q^{q''} (V + \lambda)^{-s} \d q'$ is finite for $q,\ q''$ finite),
and of $\lambda$ thanks to the regular right-hand side (at $s=-1/2$)
of the functional relation
 \begin{equation}
\label{FR}
{\partial \over \partial \lambda} \int_q^{+\infty} (V + \lambda)^{-s} \equiv
-s \int_q^{+\infty} (V + \lambda)^{-(s+1)} .
\end{equation}
That singularity is therefore an intrinsic feature of the potential $V$ alone.

The desired analytical continuation of $\int_q^{+\infty} (V + \lambda)^{-s}$ 
can be performed leftwards from $\{\Re s > 1/N\}$ using the functional relation (\ref{FR}) in reverse, 
in complete analogy with a method exposed for spectral zeta functions 
in \cite{V5}. It follows therefrom
that the only obstructions to regular continuation lie in the leading large-$\lambda$ behavior of $\int_q^{+\infty} (V + \lambda)^{-s}$, 
which decomposes into powers $\lambda^{-s+\rho}$, the integration of which
(as $\lambda^{-s+\rho+1}/(-s+\rho+1)$) yields only simple-pole singularities.
Consequently, a convenient regularization for the definition (\ref{zz}) 
allowing such a pole at $s=-1/2$ is this finite part prescription,
\begin{equation}
\label{FP}
\int_q^{+\infty} (V + \lambda)^{1/2} \defi
\lim_{s=-1/2} \left[ \int_q^{+\infty} (V + \lambda)^{-s}-{R \over s+1/2} \right]
\end{equation}
\begin{equation}
\label{RES}
{\rm with} \quad R \defi {\rm Res}_{s=-1/2} \int_q^{+\infty} (V+\lambda)^{-s}
=R(\vec v), \quad {\rm independent\ of\ } (q,\lambda).
\end{equation}
The new normalization (\ref{NN}) is then fully defined; 
{\sl it commutes with spatial translations\/}.
\smallskip

We can alternatively describe eq.(\ref{NN}) in the same framework \cite{V5}
but using $q$ as integration variable instead of $\lambda$. 
Thus, on the basis of the expansion
\begin{equation}
\label{REs}
(V(q)+\lambda)^{1/2} \sim \sum_\sigma \beta_\sigma q^\sigma \quad 
{\rm for\ } q \to +\infty, \quad
(\sigma={N \over 2},\ {N \over 2} -{1 \over N}, \cdots),
\end{equation}
\noindent $\quad \int_q^{+\infty} (V + \lambda)^{1/2}$ gets specified as 
{\sl the\/} indefinite integral consistent with the ``symbolic integration"
rules
\begin{equation}
\quad \int_q^{+\infty} (V + \lambda)^{1/2} \sim 
\sum_\sigma \beta_\sigma  \int_q^{+\infty} \! {q'}^\sigma \d q', \qquad
\int_q^{+\infty} \! {q'}^\sigma \d q' \defi 
\Bigl\{ \matrix { 
-q^{\sigma+1}/(\sigma+1) & (\sigma \ne -1) \cr 
-\log q  & (\sigma =-1) }
\end{equation}
Substitution into eq.(\ref{NN}) then yields the asymptotic behavior
\begin{equation}
\label{SUB}
\psi_\lambda (q) \sim q^{-(N/4+\beta_{-1})}
\exp \Bigl\{ -\sum_{\{\sigma>-1\}} \beta_\sigma q^{\sigma+1}/(\sigma+1) \Bigr\}
\qquad (q \to +\infty)
\end{equation}
({\sl with no outer constant prefactor\/}); 
this shows that the solution (\ref{NN}) exactly reproduces 
the subdominant solution defined in \cite{S} (ch.2, Sec.6).

\subsection{Residues, residues}

A digression is needed here to better understand the residue $R$ from eq.(\ref{RES}):
by relating various asymptotic expansions we will find an identity,
eq.(\ref{IDR}), among several independently defined constants including $R$.
\smallskip

a) We first consider the quantum partition function of the operator 
$-\d^2 / \d q^2 +V(|q|)$ on the whole line, namely
\begin{equation}
\label{TQ}
\theta(t) \defi \sum_{k=0}^{+\infty} \e^{-tE_k}.
\end{equation}
This admits a $t \downarrow 0$ asymptotic expansion obtainable as in the case
$V(q)=|q|^N$, but incorporating the lower-degree terms of $V$ as perturbations:
\begin{equation}
\label{TAS}
\theta(t) \sim \sum_\nu {\tilde c}_{-\nu} t^{-\nu}, \qquad t \downarrow 0, 
\qquad -\nu=-\mu,\ -\mu+{1 \over N},\ -\mu+{2 \over N}, \cdots
\end{equation}
(as with eq.(\ref{BS}), this expansion is taken only to some finite order here).
The leading nonclassical contribution to eq.(\ref{TAS}) again arises 
from the $|q|^N$ term for which it is known to be O$(t^\mu)$; 
therefore, below this order, eq.(\ref{TAS}) also applies to
(and is more easily computed from) 
the {\sl classical\/} partition function, i.e.,
\begin{equation}
\label{TC}
\theta_{\rm cl}(t) = \int_{\Bbb R ^2}
\! {\d p\,\d q \over 2 \pi} \e^{-(p^2+V(|q|)t} \equiv
{1 \over \sqrt{\pi t}} \int_0^{+\infty} \e^{-V(q)t} \d q .
\end{equation}

We now Mellin-transform both partition functions
(\ref{TQ}) and (\ref{TC}) term by term, as 
\begin{equation}
\label{ZQ}
{1 \over \G(s)} \int_0^{+\infty} \theta(t) \e^{-\lambda t} t^{s-1} \d t
= \sum_{k=0}^{+\infty} (E_k+ \lambda)^{-s} \defi Z(s,\lambda), 
\end{equation}
\begin{equation}
\label{ZC}
{1 \over \G(s)} \int_0^{+\infty} \theta_{\rm cl}(t) \e^{-\lambda t} t^{s-1} \d t
= {\G(s-1/2) \over \G(s) \sqrt\pi} \int_0^{+\infty} (V(q)+\lambda)^{-s+1/2} \d q
\defi Z_{\rm cl}(s,\lambda) .
\end{equation}
On the quantum side (eq.(\ref{ZQ})), this zeta function $Z(s,\lambda)$ \cite{V5}
relates to eq.(\ref{Z}) through $Z(s,0) \equiv Z(s)$,
and to eq.(\ref{DET}) through
\begin{equation}
\label{ZD}
\exp [- \partial Z(s,\lambda) / \partial s ] _{s=0} \equiv 
D(\lambda) \defi D^+(\lambda) D^-(\lambda);
\end{equation}
whereas eq.(\ref{ZC}), drawn from the rightmost eq.(\ref{TC}), 
creates a classical picture in full analogy; 
in particular, the counterpart of eq.(\ref{ZD}) evaluates using
\begin{equation}
\label{DC}
[- \partial Z_{\rm cl}(s,\lambda) / \partial s ] _{s=0} =
2 \int_0^{+\infty} (V(q)+\lambda)^{1/2} \d q 
\end{equation}
(in the sense of eq.(\ref{FP})).

(The parallelism between 
the normalization procedures (\ref{DET}) for the determinants, 
resp. (\ref{zz}) for the recessive solutions, is now fully clear:
they involve the same analytical continuations to $s=0$ of $Z(s,\lambda)$, 
resp. $Z_{\rm cl}(s,\lambda)$.)

The small-$t$ expansion (\ref{TAS}) itself Mellin-transforms to 
a large-$\lambda$ expansion
\begin{equation}
\label{ZAS}
Z(s,\lambda) \sim
\sum_\nu {\tilde c}_{-\nu} {\G(s-\nu) \over \G(s)} \lambda^{-s+\nu}
\qquad (\lambda \to +\infty)
\end{equation}
also valid strictly above O$(\lambda^{-s-\mu})$ 
for the corresponding classical function (\ref{ZC});
and the analogous treatment for the derivatives at $s=0$ \cite{V5}
likewise yields 
\begin{equation}
\label{DAS}
\log D(\lambda) \sim \sum_\nu {\tilde c}_{-\nu} [-\G(-\nu)\lambda^\nu]
\qquad {\rm with} \quad [-\G(-0)\lambda^0] \defi \log \lambda 
\end{equation}
also valid strictly above O$(\lambda^{-\mu})$ 
for the analogous classical quantity (\ref{DC}).

Below, we will specially consider the coefficient ${\tilde c}_0$;
by the preceding arguments it has a wholly classical origin, and it is:
the coefficient of $t^0$ in $\theta(t)$ or $\theta_{\rm cl}(t)$;
the coefficient of $\log \lambda$ in $\log D(\lambda)$ or 
$2 \int_0^{+\infty} (V(q)+\lambda)^{1/2} \d q$~;
the coefficient of $\lambda^{-1}$ (``residue") in 
$\int_0^{+\infty} (V(q)+\lambda)^{-1/2} \d q$
(by setting $s=1$ in eqs.(\ref{ZC}) and (\ref{ZAS}) for $Z_{\rm cl}$).

\smallskip

b) The zeros of the above determinant $D(\lambda)$ being $\{-E_k\}$, 
their asymptotic behavior is given by the Bohr--Sommerfeld formula (\ref{BS}). 
The following specifically holds before the latter begins to depend on the
parity of $k$, which may occur for $\nu \le -3/2$.
Then, the two large-energy expansions (\ref{BS}), (\ref{DAS}) 
relating to the same function have to match \cite{V1,V3}:
their exponents $\{\nu\}$ must coincide, as is already manifest,
and their coefficients must be related, actually as
\begin{equation}
\label{CMP}
{\tilde b}_\nu \equiv {\tilde c}_{-\nu} / \G(1-\nu) \qquad (\nu>-3/2).
\end{equation}
This result at once describes (the leading part of) the expansion (\ref{BS}) 
and shows that it is of classical origin up to O$(\lambda^{-\mu})$ (excluded). 
We are then going to focus on the quantity ${\tilde b}_0$ 
which contributes a constant shift term to the semiclassical series (\ref{BS}).

\smallskip

c) In the purely classical expansion (\ref{REs}) 
of $(V(q)+\lambda)^{1/2}$ for $q \to +\infty$, we select
the coefficient of $q^{-1}$, namely $\beta_{-1}$ (cf. \cite{S}, ch.2). 
By simple power counting, this coefficient is independent of $\lambda$ 
(for $N>2$) and vanishes for $N$ odd. 
For $N$ even, it is the residue at $q=\infty$ of the complex-analytic function
$(V(q)+\lambda)^{1/2} \stackrel{\rm def}{\sim} q^{(N/2)}$ (single-valued)
near $q=\infty$.
We also denote it $\beta_{-1}(\vec v)$ to stress that it is an invariant of $V$.

\smallskip

Our present goal is to obtain the following identifications, 
with $R$ as in eq.(\ref{RES}):
\begin{equation}
\label{IDR}
Z(0) \stackrel{(i)}{=} {\tilde b}_0 \stackrel{(ii)}{=} {\tilde c}_0 
\stackrel{(iii)}{=} -2R \stackrel{(iv)}{=} -{2 \over N} \beta_{-1}(\vec v)
\quad (\equiv 0 {\rm \ for\ } N {\rm \ odd}).
\end{equation}

Proof: ({\it i\/}) immediately follows from eqs.(\ref{Z},\ref{Z0}); 
({\it ii\/}), from  eq.(\ref{CMP}) specialized at $\nu=0$;
({\it iii\/}), from comparing the residues at $s=0$ of the two integrals in
eq.(\ref{ZC});
finally, we prove ({\it iv\/}) by brute force. 

On the one hand, we extract $R$ from eq.(\ref{RES}) but with $q=0$ and
integration variable rescaled by $\lambda^{-1/N}$, as
\begin{equation}
\int_0^{+\infty} (V+\lambda)^{-s} =
\lambda^{-s+1/N} \int_0^{+\infty} (x^N + 1)^{-s} \Bigl[ 1+ 
{ v_1 \lambda^{-{1 \over N}} x ^{N-1} + \cdots + v_{N-1} \lambda^{1-N \over N} x
\over x^N + 1 } \Bigr] ^{-s} \d s .
\end{equation}
The residue only affects the terms with overall weight $\lambda^{-s-1/2}$;
when the power in brackets is expanded, the corresponding coefficient comes as
\begin{equation}
\label{CO}
\sum_{k \ge 0} (-s)(-s+1) \cdots (-s-k+1) 
\sum_{\{r_j\}}{ v_1^{r_1} \cdots v_{N-1}^{r_{N-1}} \over r_1 ! \cdots r_{N-1} !}
\int_0^{+\infty} (x^N + 1)^{-s-k} x^{(N-1)r_1+ \cdots + r_{N-1}} \d x ,
\end{equation}
the inner summation being over $(N-1)$-uples $\{r_j\}$ subject to
$\sum_{j=1}^{N-1} r_j = k$ and $\sum_{j=1}^{N-1} j r_j = 1+N/2$;
hence the last written integral reduces to
\begin{equation}
\int_0^{+\infty} (x^N + 1)^{-s-k} x^{Nk-1-N/2} \d x =
{\G(s+1/2) \G(k-1/2) \over N \G(s+k)} ,
\end{equation}
which acquires the residue $1/N$ at $s=-1/2$ for any $k$.

On the other hand, we seek $\beta_{-1}$ from the large-$q$ expansion
\begin{equation}
V(q)^{1/2}= q^{N/2} 
\Bigl[ 1+ {v_1 \over q} + \cdots {v_{N-1} \over q^{N-1}} \Bigr]^{1/2} ;
\end{equation}
when the power in brackets is expanded, the coefficient of $q^{-1}$ comes out as
\begin{equation}
\sum_{k \ge 0} ({1 \over 2})({1 \over 2}+1) \cdots ({1 \over 2}-k+1) 
\sum_{\{r_j\}}{ v_1^{r_1} \cdots v_{N-1}^{r_{N-1}} \over r_1 ! \cdots r_{N-1} !}
\end{equation}
with exactly the same summation range as before, 
hence this is manifestly $N$ times the residue of eq.(\ref{CO}) 
at $s=-1/2$. \hfill QED.

\smallskip

Finally, again by simple power counting arguments, 
we remark that these classical invariants are not only absent for odd $N$, 
but also vanish for another broad class of potentials:
all purely even polynomials of degree $N$ multiple of 4 
(as well as for all $N$ in the special homogeneous case: 
$\beta_{-1}(\vec 0) \equiv 0$.)
Barring $N=2$, nontrivial residues then first occur 
for non-even quartic potentials:
\begin{equation}
\beta_{-1}(\vec v) = {v_3 \over 2} -{ v_1 v_2 \over 4} + {v_1^3 \over 16}
\qquad (V(q) = q^4 + v_1 q^3 + v_2 q^2 + v_3 q),
\end{equation}
and, within even potentials, for sextic ones:
\begin{equation}
\beta_{-1}(\vec v) = {v_4 \over 2} -{ v_2^2 \over 8}
\qquad (V(q) = q^6 + v_2 q^4 + v_4 q^2).
\end{equation}

\subsection{Basic identities}

Under the above notations, very simple identities connect
the spectral determinants and the absolute-normalized solution:
\begin{equation}
\label{ID}
D^-(\lambda ) \equiv \psi_\lambda (0), \qquad 
D^+(\lambda ) \equiv -\psi'_\lambda (0).
\end{equation}
(The proof is an adaptation of the arguments in \cite{V1}, Apps.~A and D).

Next, following \cite{S}, we continue eq.(\ref{Schr}) in the complex $q$-plane
down to the rotated half-line lying in the adjacent Stokes direction, namely
$ [0,\e^{-\mi\varphi/2} \infty) $ where
\begin{equation}
\varphi \defi {4 \pi \over N+2} \quad  (\mbox{\sl spectral symmetry angle}).
\end{equation}
By simple complex scaling upon $q$,
\begin{equation}
\psi^{[1]} \defi \psi_{ \e^{-\mi\varphi} \lambda} 
(\e^{\mi\varphi/2}q ; {\vec v}^{[1]})
\end{equation}
provides another solution (to eq.(\ref{Schr})) now recessive
in the Stokes direction $\e^{-\mi\varphi/2}$, where
\begin{equation}
{\vec v}^{[1]} \defi (\e^{\mi\varphi/2} v_1, \e^{2\mi\varphi/2} v_2, \cdots,
\e^{(N-1)\mi\varphi/2} v_{N-1})
\end{equation}
expresses an action of the discrete rotation group of order $(N+2)$ on the coefficients; equivalently it acts upon the potential $V$, mapping it to
$V^{[1]}$, then $V^{[2]}, \cdots$ (now {\sl complex\/} potentials). 
The order of the {\sl effective\/} symmetry group is 
\begin{equation}
\label{ELL}
L=N+2 \mbox{ generically}, \qquad L=N/2+1 \mbox{ for an even polynomial } V.
\end{equation}

Now the Wronskian of the two solutions $\psi, \psi^{[1]}$ of eq.(\ref{Schr}), 
a constant, can be evaluated explicitly from their respective asymptotic forms
(\ref{SUB}), both valid as $q \to +\infty$ \cite{S},
and also expressed at $q=0$ by means of the respective identities (\ref{ID})
for the potentials $V$ and $V^{[1]}$. 
Matching the two calculations then yields the fundamental bilinear identity
\begin{equation}
\label{DW}
\e^{+\mi\varphi/4} D^+( \e^{-\mi\varphi} \lambda, {\vec v}^{[1]} )
D^-( \lambda, {\vec v})
-\e^{-\mi\varphi/4} D^+( \lambda, {\vec v}) 
D^-( \e^{-\mi\varphi} \lambda, {\vec v}^{[1]} ) \equiv 
2 \mi \e^{\mi\varphi\beta_{-1}(\vec v) /2}.
\end{equation}

We stress that our approach will bypass any other matching of solutions,
like those required in connection problems between nonadjacent 
Stokes directions, which yield nontrivial Stokes multipliers...

Remarks: a) in the harmonic case ($V(q)=q^2,\ N=2,\ \varphi=\pi$), 
many of our arguments become invalid but the end result (\ref{DW}) remains,
except that the residue is now $\lambda$-dependent,
$\beta_{-1}=\lambda/2$ from eq.(\ref{REs}) (the identity (\ref{DW})
now verifies by applying the reflection formula for $\G(z)$ to
$D^+(\lambda) = 2 \sqrt{\pi}\ 2^{-\lambda/2} / \G ({1+\lambda \over 4})$, 
$D^-(\lambda) = \sqrt{\pi}\ 2^{-\lambda/2} / \G ({3+\lambda \over 4})$); 
b) identities similar to eq.(\ref{DW}) have also surfaced 
in quantum integrable theories, as ``quantum Wronskian conditions" \cite{QW}.

\section{Exact quantization conditions}

For a homogeneous potential, we currently believe that the functional relation
(\ref{DW}) and the asymptotic law (\ref{BS}) (imposed to some o(1) accuracy) 
together suffice to specify the whole spectrum exactly:
indeed, we empirically recovered the spectrum as the fixed point of 
an (apparently) contractive mapping built using just that input \cite{V2,V3,V4}.
Thus, for zeta-regularized products built over sequences with prescribed
Bohr--Sommerfeld asymptotics, the single Wronskian identity (\ref{DW}) 
can become so coercitive as to fix both its arguments completely.
We now make the same guiding idea work for general polynomial potentials.

\subsection{The analytical result}

Guided by the homogeneous case \cite{V3}, we strive to turn eq.(\ref{DW}) 
into equations where at least the Neumann ($+$) and Dirichlet ($-$) spectra
appear decoupled.
To keep, say, only the former (the other admits a mirror-image treatment), 
we take eq.(\ref{DW}) and its partner with ${\vec v}^{[-1]}$
(written invoking the homogeneity property
$ \beta_{-1}({\vec v}^{[\ell]}) \equiv (-1)^\ell  \beta_{-1}(\vec v) $),
then we set $\lambda=-E_{2k}$ and eliminate $D^-( \lambda, {\vec v})$
from the resulting pair, to find
\begin{equation}
\label{MUL}
D^+(\e^{-\mi\varphi}\lambda,{\vec v}^{[1]})
/ D^+(\e^{+\mi\varphi}\lambda,{\vec v}^{[-1]})
\vert_{\lambda=-E_{2k}} = -\e^{-\mi\varphi/2\,+\, \mi\varphi\beta_{-1}(\vec v)}.
\end{equation}
Whereas the procedure for $V(q)=q^N$áclosed upon itself at once, 
this general one invokes the complex-rotated potentials $V^{[\pm 1]}$ and,
step by step, all the $L$ partner potentials $V^{[\ell]}$ to reach closure.
The spectra $\{E_{2k}^{[\ell]}\}$ (mostly complex) 
then altogether make up the independent unknowns 
(they are not independent as analytical continuations of each other in complex
${\vec v}$-space, but we are unable to get explicit relations expressing this). 

Next, absolute phases are specified in eq.(\ref{MUL}) 
(and its cyclic permutations on the ${\vec v}^{[\ell]}$)
by reference to the homogeneous case \cite{V3}.
This operation is essential to create quantization relations
in the Bohr--Sommerfeld form
(governed by an explicit quantum number $k=0,1,2,\ldots$), but now exact. 
Our final result, in a synthetic notation, is an uncoupled pair of systems
(using either ($+,\ k$ even) for Neumann, or ($-,\ k$ odd) for Dirichlet), 
each built of $L$ exact quantization conditions: one per potential $V^{[\ell]}$ 
(with $\ell$ integer mod $L$),
\begin{eqnarray}
\label{EQC}
\Sigma_\pm^{[\ell]}(E_{k}^{[\ell]}) &=& 
\pi \Bigl[ k+{1 \over 2} \pm {N-2 \over 2(N+2)} \Bigr]
+ (-1)^\ell \varphi \beta_{-1}(\vec v) \qquad 
{\rm for\ } k= 
{\scriptstyle{\scriptstyle{0,2,4,\ldots}\atop \scriptstyle{1,3,5,\ldots}}}
\nonumber\\
{\rm where} \quad \Sigma_\pm^{[\ell]}(E) &\defi& 
-\mi \Bigl[ \log D^\pm(-\e^{-\mi\varphi}E,{\vec v}^{[\ell+1]})
-\log D^\pm(-\e^{+\mi\varphi}E,{\vec v}^{[\ell-1]}) \Bigr] , 
\end{eqnarray}
the branches of $\log D^\pm(\lambda)$ being taken by continuity from 
$(\lambda=0,\ {\vec v}=0)$. 

The harmonic case $V(q)=q^2$ (with $N=L=2,\ \varphi = \pi$) holds again with a
degenerate structure, $\Sigma_\pm^{[\ell]}(E) \equiv 0$ but $\beta_{-1} = -E/2$.

\subsection{Discussion}

All following considerations are meant in either spectral sector independently 
(Neumann, resp. Dirichlet), 
with all quantum numbers accordingly kept even, resp. odd.

Eqs.(\ref{EQC}) form a system of constraints 
tying each $E_{k}^{[\ell]}$ at fixed $\ell$ 
to the two other spectra 
$\{E_{j}^{[\ell-1]}\}$, $\{E_{j}^{[\ell+1]}\}$ 
(whose zeta-regularized products build the determinants defining 
$\Sigma_\pm^{[\ell]}$). 
The system of all such points in interaction is better displayed,
in proper relative positions, if each spectrum is suitably rotated
(we then call it a `chain', as a more general name): 
for $\ell=0,\ldots,L-1$ (mod $L$), the $\ell^{\rm th}$ such chain is the set
$ \{\e^{\mi\ell\varphi} E_{k}^{[\ell]}\} $, 
and each of its points is under the influence of every point 
in the two adjacent ($(\ell \pm 1)^{\rm th}$) chains --- through the logarithm
of their complex difference which enters eq.(\ref{EQC}) via the formula (\ref{EMc}).
(In the generic case $L=N+2$, adjacency is correctly shown on a double covering of the circle of asymptotic directions; the circle itself suffices only in the purely even case $L=N/2+1$.)

The $\ell^{\rm th}$ equation is now a complex one precisely when its unknowns 
$E_{k}^{[\ell]}$ are themselves complex, i.e., for $\ell \ne 0$ or $L/2$:
eq.(\ref{EQC}) thus remains a formally `complete' system of mutual constraints
for the unknowns $E_{k}^{[\ell]}$.
As in the homogeneous case we then surmise that eqs.(\ref{EQC}) 
are not only exact, but also genuinely complete, quantization conditions; 
i.e., they have the capacity to determine all their unknowns 
provided the asymptotic condition (\ref{BS}) is also enforced 
upon each spectrum $\{E_{k}^{[\ell]}\}$ separately 
(using the rotated coefficients ${\vec v}^{[\ell]}$). 
If moreover this resolution can be performed in any constructive way,
then we may argue that the analytical formulae (\ref{EQC}) themselves are  
``giving" the solution of the stated spectral problem
(Dirichlet or Neumann on the half-line).

We will now report numerical experiments, mostly upon quartic potentials, 
which empirically confirm our conjecture in some regions of parameter space,
by achieving {\sl effective\/} computations of the spectra 
out of eqs.(\ref{EQC}) (plus eqs.(\ref{BS})). 

\section{Numerical tests of exact spectrum quantization}

At present we can further resolve eqs.(\ref{EQC}) only numerically.
We will essentially seek to solve them by successive approximations 
for each level $E_{k}^{[\ell]}$ in turn (by Newton's root-searching method), 
but stopping at some finite $k_{\rm max}$, beyond which 
all eigenvalues can assume semiclassical values instead, once for all.
The height of the cutoff $k_{\rm max}$, together with the depth in use for
the semiclassical expansion (\ref{BS}), control the final accuracy.
As for the zeta-regularized products involved, we use formula (\ref{EMc})
(again adding higher expansion terms to improve the $k_{\rm max}\to \infty$
convergence).
\smallskip

The preceding formalism immediately provides the quantization of levels 
for an even polynomial potential on the real line: parity symmetry
splits the spectrum into even and odd sectors,  which exactly correspond
to the Neumann and Dirichlet problem on the half-line, respectively.
As a side effect, parity also halves the order of symmetry $L$ 
(cf. eq.(\ref{ELL})).

\subsection{Even quartic oscillators}

We now test the exact framework upon the quantization of levels 
for even potentials $V(q)=q^4+v_2 q^2$ on the real line
(the order of symmetry being $L=3$).

To the initially real $v_2$ giving the real spectrum $\{E_k\}$ 
are then associated:
a complex spectrum $\{E'_k\}$ for the coupling constant $\j v_2$,
and $\{E''_k\}$ for $\j^2 v_2$ (where $\j=\e^{2\pi\mi/3}$). 
The three chains $\{E_k\},\ \j\{E'_k\},\ \j^2\{E''_k\}$ are shown on fig.1
for selected values of the harmonic coupling constant $v_2$.
They altogether constitute the reduced dynamical unknowns;
by reality symmetry, there are only two independent chains
(one real, and one complex).

Then, in order at once to reach an equilibrium point for all constraints and 
to gain evidence for its uniqueness, we iteratively apply an elementary step: 
to recompute each chain in turn as (numerical) solution of its eq.(\ref{EQC})
in terms of the adjacent chains at their current locations.
Through an appropriate succession of such steps, 
we then try to build up a global process yielding a contractive iteration,
as in the homogeneous case 
(now many more fine details can vary, and we tried but a few combinations). 
For us, a numerical validation of the exact quantization formulae results 
if one such iteration is found which displays
{\sl geometric\/} convergence to the {\sl correct\/} spectra, i.e.,
a contraction ratio can be estimated with a minimum of numerical stability,
{\sl and\/} the limiting chains (checked upon their lowest five points, say)
agree with independent eigenvalue calculations (e.g., matrix diagonalization). 

In order to define the mandatory asymptotic behaviors of all chains, 
we also need reference semiclassical chains for all three rotations of $v_2$.
These we take as the (numerical) solutions of eq.(\ref{BS})
pushed to 6 terms (=O($E^{-1/2}$), for better $k \to +\infty$ convergence).
The same chains also conveniently serve as initial iteration data.

Such an iterative approach was found for the homogeneous case ($v_2=0$) 
and it yielded strong contraction ratios ($\losim 0.4$) \cite{V2,V3}. 
However, this case also enjoyed full ternary rotation symmetry,
and a qualitative change occurs now that this symmetry has to be relaxed.
If we try to deform those earlier $v_2=0$ iterations 
into $v_2$-dependent schemes, the latter must apply synchronous updating
to keep with the full ternary symmetry at $v_2=0$ 
(i.e., the chains are recomputed individually but 
get actually updated all at once, at each completion of a full $\ell$-cycle).
Unfortunately, such iteration schemes also appear to
excite new (symmetry-breaking) chain fluctuation modes 
possessing much lower, almost marginal, stability 
(contraction ratios $\approx \pm 0.9$, down to $v_2 = 0$). 
As a consequence, their convergence discontinuously drops 
in the immediate vicinity of $v_2=0$, down to unacceptably low levels.
On the other hand, we empirically found that iteration schemes of a different
type, which {\sl immediately\/} update the chains one by one,
can remain fairly stable in a finite neighborhood of $v_2=0$
(even though they are not as good as the earlier ones at the value $v_2=0$,
where now the ternary symmetry is spontaneously broken).

The simplest such scheme (``A") consists of
recomputing the real and the complex chain alternatively.
For $v_2$ near 0, this already appears to converge nicely 
and it indeed yields a first validation of the exact formalism. 
Still, its behavior deteriorates as $v_2$ grows:
between $v_2 \approx +2$ to $+3$, 
the iterations start to converge more erratically, 
then our root-searching algorithm (the Newton method) goes unstable, 
and shortly afterwards the contraction ratios
of the iteration (plotted on fig.2, left) shoot up to almost unity
(for $v_2 < 2$ they were estimated both from the iteration sequence and 
from diagonalizing the linearized dynamics near the fixed point, 
and for $v_2 \ge 2$ only the latter way).
In particular, scheme ``A" cannot be carried towards the harmonic limit. 
By contrast, it behaves much better in the negative direction 
(double-well region), showing no sign of degradation over our test range,
i.e., down to  $v_2 \approx -10$~!

The positive-$v_2$ breakdown is possibly explained by fig.1 (right): 
when $v_2$ grows, the complex chain and its conjugate become almost degenerate
towards low quantum numbers (a sort of complex tunneling effect), 
causing larger logarithmic pair interactions 
and instabilities (with the linearized-dynamics
matrix entries growing like [mutual distance]$^{-1}$).
An obvious remedy is then to remove the interaction between any two
complex-conjugate chain points from the dynamics, by enforcing this symmetry
as an {\sl a priori\/} constraint (immediate updating also has to be kept);
the so modified scheme (``B") exhibits a much more uniform linear stability
indeed (fig.2, right). 
This scheme converges and allows to validate the exact formalism for larger
values of $v_2$, up to $v_2 \approx +5$.
Beyond, we again become unable to get any convergence;
so, global instabilities must still be creeping in, albeit more slowly.
Again, the root-searching algorithm (the Newton method) diverges first,
but now the linear contraction ratios themselves hardly grow at all
(as obtained by diagonalization).
Hence scheme ``B" seems not to break down like ``A" 
but rather to reach certain practical limits, 
beyond which only more elaborate implementations might resolve the case.
(Perhaps, e.g., a more robust root-searching method would work, 
or the naive branch prescriptions implied in eq.(\ref{EQC}) 
ought to be revised at such a distance from the homogeneous case.)

The preceding break-points referred to iterations in the even-parity sector.
As fig.2 shows, the odd sector tends to behave more stably.
A conceivably better idea might thus be to confine iterations 
to the odd spectrum, then to solve for the even determinants 
from the coupling relations (\ref{DW}) instead.

In conclusion, the preceding tests favor the validity and effectiveness of
the exact quantization mechanism for even quartic potentials over a sizable
range of the harmonic coupling $v_2$ about 0, 
but are currently inconclusive for $v_2 \grsim +5$.

In addition, we lately extended the scheme to sextic even potentials, 
with qualitatively similar results for the (still few) cases tested.

\subsection{Extension to non-even potentials}

In \cite{V4} we showed that the exact quantization formalism was fully valid 
for homogeneous potentials of any odd degree $N$, and specially for $N=1$ 
which has Airy functions both as solutions and as spectral determinants. 
The paradox in the very regular behavior of this case is that 
the underlying potential $V(q)=|q|$ is not even once differentiable at $q=0$.
This example paved the way to the present generalization, establishing that 
the `even/odd' decomposition needed in the homogeneous case
was just a Neumann/Dirichlet splitting,
any parity properties of the {\sl polynomial\/} $V$ being irrelevant: 
an even potential over the whole line is always present as $V(|q|)$
(once its continuous differentiability is recognized as immaterial).

We thus also tested the exact quantization conditions (\ref{EQC}) 
upon a few non-even quartic polynomials (now for the Dirichlet/Neumann spectra).
The exact quantization conditions then involve $L=6$ distinct chains: 
$\ell=0,\ 3$ are real, $(1,5),\ (2,4)$ being doublets of complex conjugates. 
We found one iteration sequence (``C") to converge better than others 
(for no clear reason):
$\ell=\{0,2,3,1\}$ (then cyclically continued, with immediate updating 
of each chain and straight enforcement of 1--5 and 2--4 symmetry).
The overall results were then comparable to those above for even polynomials 
(though occasionally achieving a lower numerical accuracy). 

This extension of the exact quantization formalism 
to non-even polynomials crucially opens the way to another application 
(and validation) now to be described.

\section{Exact wave-function analysis}

\subsection{Analytical reconstruction of the wave function}

If no parity property is imposed upon the polynomials $V(q)$ 
with respect to the endpoint $q=0$,
then this endpoint can be taken to an arbitrary value $a$ 
(on the real line, at least), 
thus restoring translational invariance effectively. 
Equivalently, 0 is kept as endpoint but the potential gets shifted to 
$V_a (q) \defi V(q+a)-V(a)$ on the half-line (becoming $V(|q|+a)-V(a)$ on the whole line). 

So, we now use eq.(\ref{ID}) at $q=a$ instead of 0 and from right to left, 
to state that the absolute-normalized solution of eq.(\ref{Schr}) 
at $q=a$, resp. its first derivative, are specified as
\begin{equation}
\label{SOL}
\psi_\lambda (a) \equiv D_a^-(V(a)+\lambda), \qquad 
\psi'_\lambda(a) \equiv -D_a^+(V(a)+\lambda),
\end{equation}
in terms of the spectral determinants for the modified potential $V_a (q)$ 
on $[0,+\infty)$, denoted $D_a^\pm$. 
But these in turn are canonically specified as 
the zeta-regularized products over their own chains of zeros; now the latter, 
as the eigenvalues of $(V_a) ^{[\ell]}(q)$, are given by 
exact quantization conditions of the form (\ref{EQC}),
hopefully through some convergent iteration scheme as above.
Thereupon, eq.(\ref{SOL}) simply asks to apply the zeta-regularized 
product formula once more (at one prescribed point, $(V(a)+\lambda)$),
in order to output the absolute-normalized solution $\psi(a)$ (or $\psi'(a)$).
Thus, eq.(\ref{SOL}) acts as the last instruction in a procedure 
to solve the full differential equation (\ref{Schr}) 
--- for arbitrary $\lambda$ --- through eqs.(\ref{EQC})
(still subject to the latter giving convergent iterations,
if an effective algorithm is wanted).

As a further prospect, quite general spectral problems 
can subsequently be tackled in principle. 
E.g., to find the eigenvalues $\lambda$ of eq.(\ref{Schr}) on the whole line
for a general potential of even degree,
one may proceed to solve eq.(\ref{Schr}) as just explained 
but from both ends of the $q$-axis, 
then match the two resulting values for 
$(\psi'/\psi)(a)$ at some finite location $a$.

\subsection{A numerical test}

We now validate eq.(\ref{SOL}) with a calculation of 
the ground state eigenfunction for the homogeneous case $V(q)=q^4$. 
The shifted potential is then $ V_a(q) = q^4 + 4 a q^3 + 6 a^2 q^2 + 4 a^3 q $,
and $\psi(a)$ is the value of its determinant $D^-(\lambda)$ at the point $\lambda = a^4-E_0$,
with the eigenvalue $E_0 \approx 1.06036209$ being part of the input here
(while it belongs to the output of an exact level calculation for $V(q)=q^4$
itself \cite{V2}).

The exact quantization conditions for $V_a(q)$ are involving 
$L=6$ distinct chains (except at $a=0$, where $L=3$ by parity symmetry).
We only tested calculations of $\psi(a)$ (not $\psi'(a)$)
for a few values of $a \ge 0$,
and found the above iteration scheme ``C" to converge indeed for 
$a \losim 1.7$ (with a contraction ratio per cycle $ \losim 0.67$).
Instabilities made us unable to pin down convergent iterations
for $a \grsim 1.7$ 
(similar comments apply as for scheme ``B" when $v_2 \grsim +5$).
Our output points are plotted on fig.3, against the curve produced by 
a standard integration routine and upon which
only the global normalization was fitted ({\sl a posteriori\/}): 
the results show an overall 4--5-digit agreement.

\subsection{Concluding remarks}

We have reduced the resolution of polynomial Schr\"odinger equations
(\ref{Schr}) to that of a discrete system of selfconsistent exact quantization conditions,
eqs.(\ref{EQC}), having as unknowns $(N+2)$ countable sequences of points 
subject to asymptotic boundary conditions consisting of
standard Bohr--Sommerfeld formulae (\ref{BS}).
Eqs.(\ref{EQC}) are supplying explicit equilibrium conditions for those
semi-infinite and asymptotically tethered chains.

We also have growing {\sl numerical\/} evidence that the so reduced problem 
is effectively solvable in some regions of parameter space
through iterations which seem to converge geometrically:
it thus appears to be a self-stabilizing system, 
in sharp contrast with the original Schr\"odinger dynamics.
This strengthens our following conjecture: (in such cases) 
the relevant equilibrium solution is realized as a fixed point of a 
contractive mapping, which also admits robust finite-dimensional approximations.
This results in an overall indirect constructive mechanism, 
where exact quantization formulae explicitly specify the mapping only.

On the darker side, since eqs.(\ref{EQC}) are very tied to the values of the
degree $N$ and symmetry order $L$, they may be ill-suited to transitional
regions where one of these numbers jumps 
(e.g, $v_2 \to\ 0\ {\rm or\ }+\infty$ in $V(q)=q^4 + v_2 q^2$).
Moreover, our tests still span a limited range,
mostly quartic potentials close to $q^4$, a few even sextic ones, 
and (previously \cite{V3}), homogeneous potentials for their spectra only 
(but up to quite high degrees). Effectiveness for higher $N$, 
and extensions to arbitrary complex $q$ and $\vec v$, 
to more general differential systems, etc., 
are all conceivable but they remain open issues.

Finally we argue that, while the integration of eq.(\ref{Schr}) 
may still be a remote goal using quadratures alone,
another valuable question is how much the set of admissible integration methods needs to be enlarged to reach that same purpose. 
Our findings provide clues to the latter issue: they strongly suggest that
zeta-regularized infinite products (of order $<1$) 
plus the solving of one type of fixed-point equations 
(which seem to have contractive and other nice properties) are
pertinent additional integration tools, 
which are possibly sufficient to the task
(already in several cases, at least).

\medskip

{\bf Acknowledgment:} we are grateful to R. Guida (Saclay) for letting us
use his computer program to calculate hundreds of anharmonic oscillator levels with high accuracy.

\vfill\eject

\centerline{\bf Figure captions}
\bigskip

Fig.~1. Interacting chains for even quartic potentials $V(q) = q^4 + v_2 q^2$,
shown for odd spectral sector (i.e., odd quantum number $k$ throughout).
The chains occupy their equilibrium positions $\{ \j^\ell E^{[\ell]}_k\}$
(points marked with the corresponding $\ell$-value).
The homogeneous case $v_2 =0$ (at center) 
has full (ternary plus complex-conjugation) symmetry.
As $v_2 \to -\infty$, the complex chains tend to shadow 
the resonance spectrum $\{\pm (2k+1) \mi \sqrt{-v_2} \}$
of the potential $-|v_2| q^2$;
as $v_2 \to +\infty$, they tend to shadow the sequence
$\{-(2k+1)\sqrt{v_2} \}$
($\ell=1$ chain for the potential $v_2 q^2$).

\medskip

Fig.~2. Linear contraction ratios for iteration schemes associated 
with potentials 
$V(q) = q^4 + v_2 q^2$ ($+$: even spectrum; $\circ$: odd spectrum).
Left: numerical estimates for iteration scheme ``A" 
with immediate updating (see main text);
although this remark is inconclusive, we add
that the numerical values of the ratios remain strictly below unity
(no error estimate at all is implied in our data,
but 2--3 digit stability is typically seen).
Right: numerically estimated {\sl moduli\/} of the ratios for
iteration scheme ``B" which further decouples conjugate-pair interactions
(here, various parts of data may correspond to different eigenvalue branches, some being negative or in complex pairs). 

\medskip

Fig.~3. Calculations of the ground-state eigenfunction
for the homogeneous quartic potential $q^4$. 
$+$: absolute-normalized data points obtained by iterative exact quantization
(scheme ``C"); in superposition,
$\diamond$: numerical estimates for the corresponding contraction ratios. 
Curve: 
computer integration of the Schr\"odinger equation by the NAG routine D02KEF;
its rescaling factor was the only number fixed by a fit.

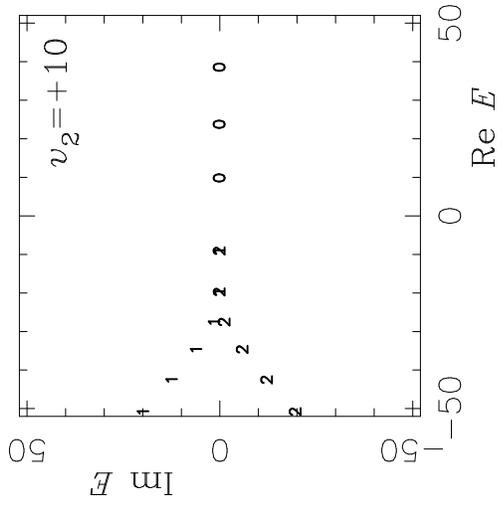
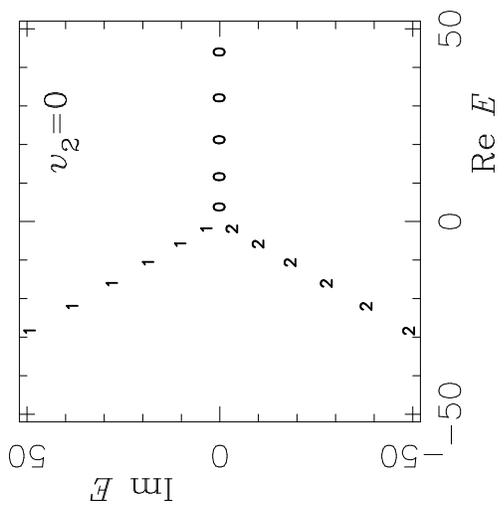
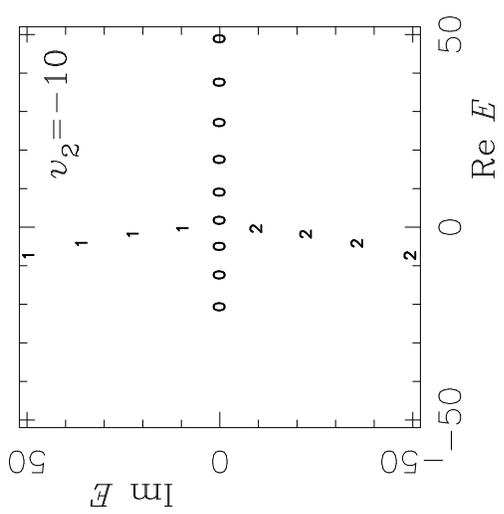
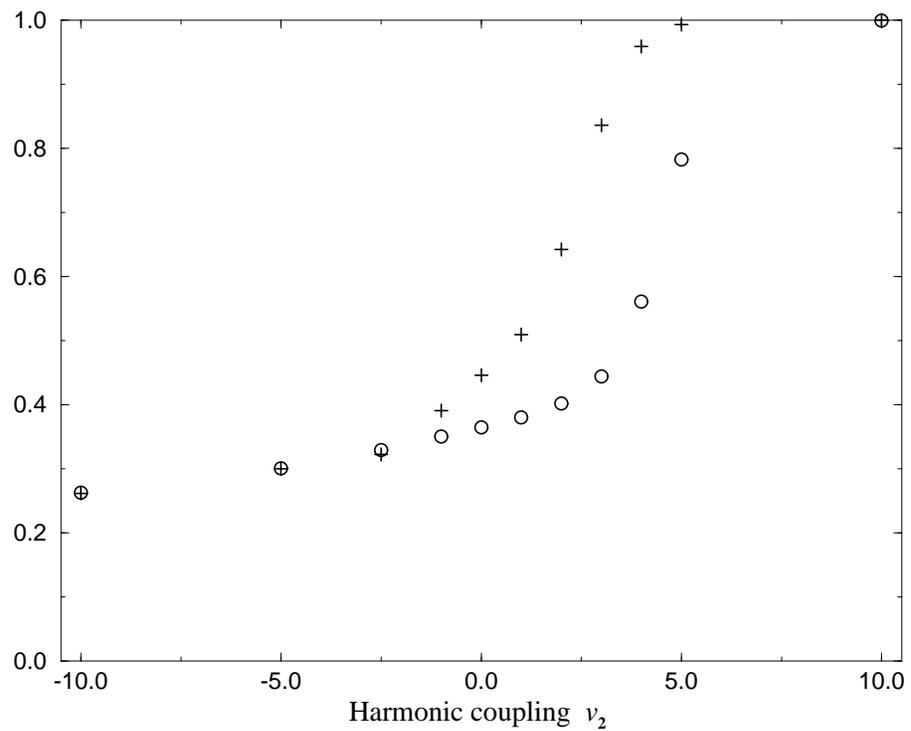
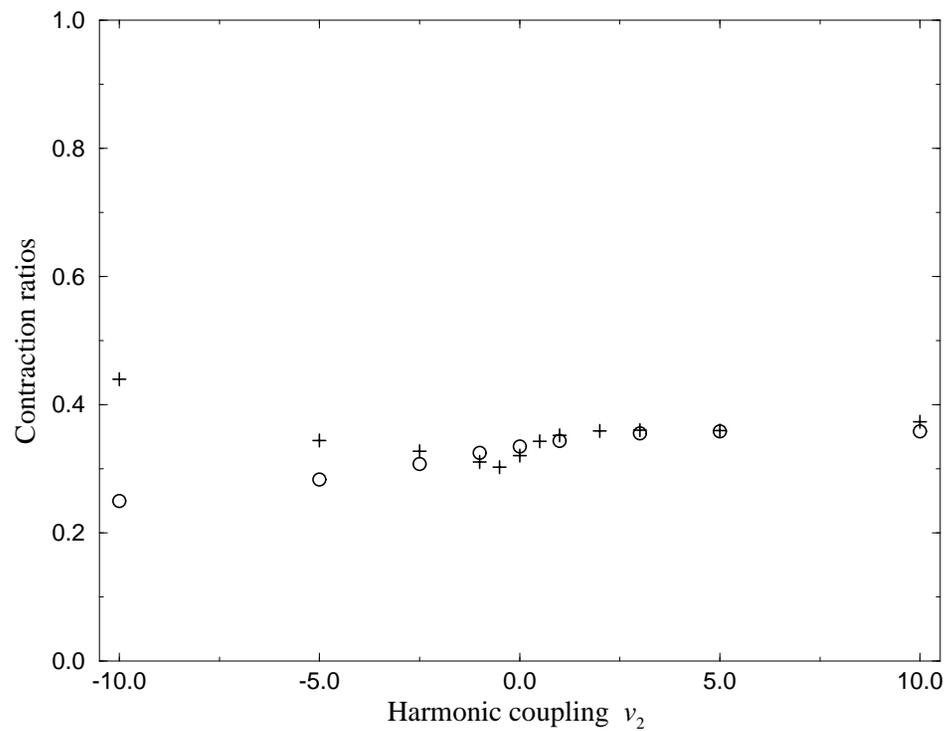
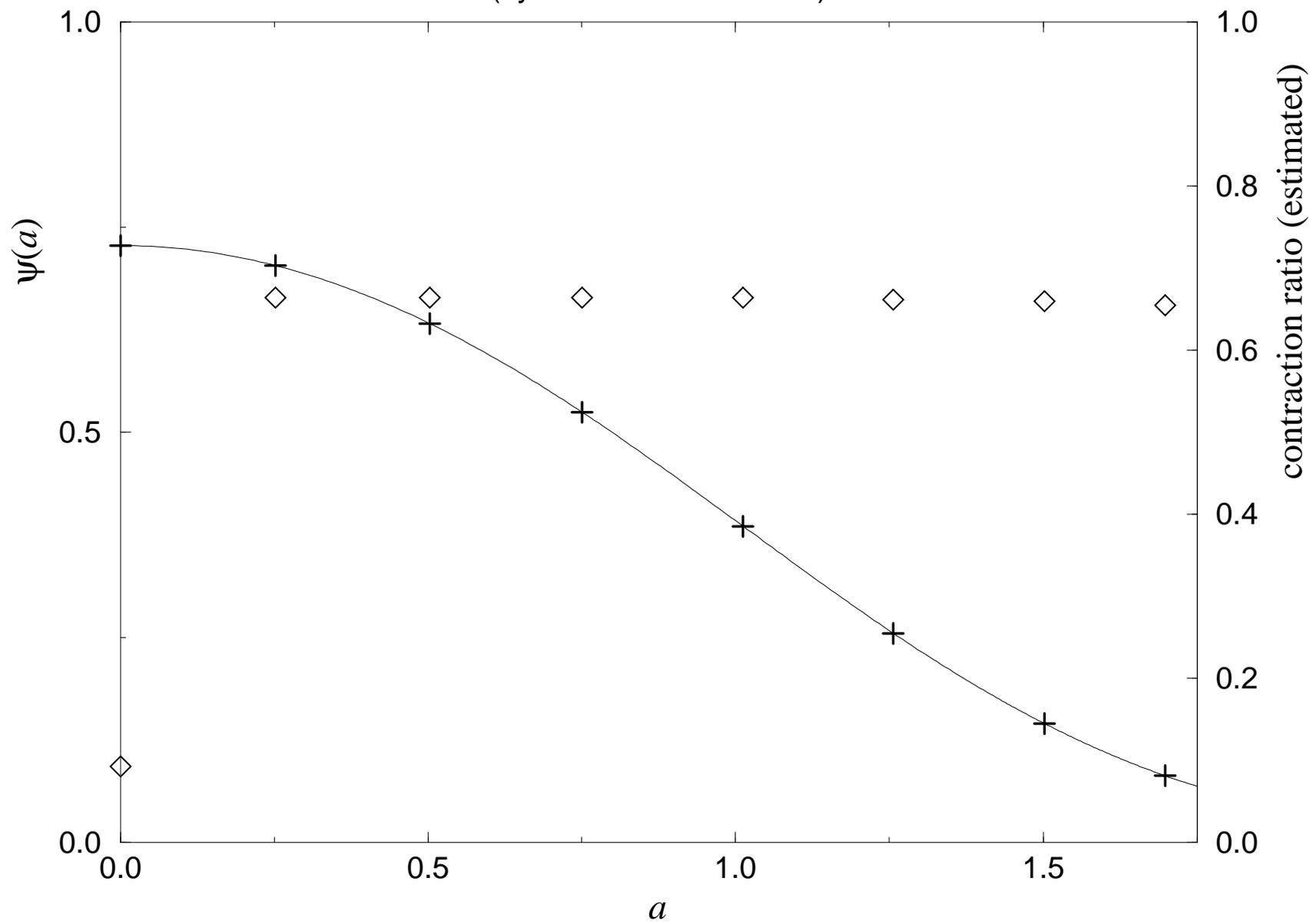
\end{document}